\documentclass[
&preprint,
longbibliography,
amsmath,amssymb,
aps,
prb,
twocolumn
]{revtex4-2}
\usepackage{graphicx}
\usepackage{dcolumn}
\usepackage{bm}
\usepackage{epsfig}
\usepackage{afterpage}
\usepackage{lineno}

\begin{document}

\title{Criterion for Vestigial Order above a Nematic Superconductor }

\author{P. T. How}
\email{pthow@outlook.com}
\affiliation{Institute of Physics, Academia Sinica, Taipei 115, Taiwan}
\author{S. K. Yip}
\email{yip@phys.sinica.edu.tw}
\affiliation{Institute of Physics, Academia Sinica, Taipei 115, Taiwan}
\affiliation{Institute of Atomic and Molecular Sciences, Academia Sinica, Taipei 115, Taiwan}

\date{\today}

\begin{abstract}

A nematic superconductor can in principle support a vestigial order phase above its superconducting transition temperature,
 with rotational symmetry spontaneously broken while remain non-superconducting.
We examine the condition for this vestigial nematic order to occur, within a Ginzburg-Landau theory with order parameter fluctuations included. 
Contrary to prior theoretical results, we found that this vestigial order actually requires very stringent conditions to be met:
the material must be sufficiently deep in the nematic regime (i.e. far away from the boundary separating the nematic and chiral superconducting phases) to  possibly exhibit a vestigial nematic order.


\end{abstract}

\maketitle



\section{Introduction}

   Superconductivity in doped topological insulator Bi$_2$Se$_3$ has captured much recent attention.   While the crystal
is supposed to have D$_{3d}$ symmetry, it has been found experimentally that the NMR Knight shifts \cite{NMR} and
the upper critical fields \cite{deVisser}  have two-fold anisotropy in the basal plane.  These are explained by the proposal that
superconductivity in this system is {\em nematic} \cite{Fu14}.   More precisely, it has been proposed
that the superconducting order parameter belongs to a two-dimensional representation, and
the energetics is such that, below the superconducting transition, the order parameter picks
a state with spontaneously broken rotational symmetry (other than the other possibility 
where time reversal symmetry is broken, {\it c.f.} the case for UPt$_3$ \cite{Sauls,Joynt}). 
Two-fold symmetry
breakings have been observed also in many other experiments, as reviewed in \cite{Yonezawa}. 

If the order parameter belongs to a two-dimensional representation, one expects an internal degree of freedom,
in this case, rotation of the order parameter, to reveal itself under suitable circumstances.
 However, so far no experiments have convincingly shown this degree of freedom.
One may expect external stress can re-orient the order parameter \cite{Nem}, but
an experiment at Argonne \cite{Willa} turns out to be negative. 
In a related experiment on multidomain sample at Kyoto \cite{Kostylev},
 only changes of the relative sizes of the domains were found.   One might also expect that
there should be special features in the upper critical field such as kinks as a function of the magnitude of the field
\cite{Willa18} ({\it c.f.} \cite{Hess}) or angle in the plane \cite{Vanderbos16}.
Neither has been reported so far and a recent experiment \cite{Bannikov} specifically looking for these features
was not able to find one.   Other theoretical suggestions have also been made in the literature.
Others \cite{Zyuzin} and us \cite{HQV} have predicted the existence of half quantum vortices or skyrmions (which
are unique to multicomponent order parameters but absent in single component systems).
 We have also investigated the special features in shear stress tensor
due to the multi-dimensional nature of order parameter \cite{Shear}.   Experiments examining these predictions have
not yet been reported.

A nematic superconducting order parameter breaks both gauge and rotational symmetry.
In principle these two broken symmetries do not necessarily occur at the same temperature. 
A few years ago,  \cite{Hecker18} predicted that ``vestigial nematic order" can exist in this
system:  as the temperature is lowered, the symmetry preserving normal state first makes
a transition into a state with broken rotational symmetry, and only later gauge symmetry
is broken, forming the nematic superconducting state.  This possibility is unique to
a multi-component order parameter:  a superconductor with an order parameter
belonging to a one-dimensional representation, even if it is not s-wave, cannot 
exhibit this vestigial order.  Observation of this ``vestigial nematic state" would
be a ``smoking gun" of this nature of the order parameter.   Not long after this proposal,
an experiment \cite{Cho} indeed claimed that this vestigial order has been observed. In particular,
   length change of the sample as a function of temperature or field was monitored. 
A rapid and directional dependent change as a function of temperature  above the superconducting transition was observed and interpreted
as a step indicating a first order transition into a vestigial nematic order state.
It is remarkable that the relative change in length is only of order $10^{-7}$, even smaller
than the distortion from perfect D$_{3d}$  found at higher temperatures from another group \cite{Kuntsevich}.
Vestigial orders have been recently discussed in many other systems \cite{VReview,Grinenko21}

\begin{figure}[h]
\includegraphics[width=0.5\textwidth]{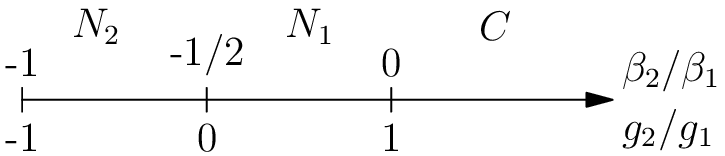}
\caption{
The different regions for the quartic coefficients.  ($\beta_1 > 0$). Within mean field, the superconducting ground state would be nematic in both regions N$_1$ amd N$_2$, but chiral in region $C$.
In this paper, we find that vestigial nematic order above the superconducting transition occurs only in region N$_2$ but not N$_1$.}
\label{figp}
\end{figure}

In a Ginzburg-Landau formulation, superconducting order parameter belonging to a two-dimensional representation
in a D$_{3d}$ system has two ``interacting" constants, or coefficients entering the quartic terms of the free energy.
(e.g. $\beta_{1,2}$ in our notations (\ref{Hint}) below).  These parameters dictate whether the
mean-field superconducting ground state of the system would have nematic order (in our case $ - \beta_1<\beta_2 < 0$)
 or broken time-reversal symmetry ($\beta_2 > 0$).  
Ref. \cite{Hecker18}, analyzing the problem using a Hubbard-Stratanovich transformation,
 concluded that all regions with $ - \beta_1< \beta_2 < 0$  with a nematic superconducting ground state
can potentially exhibit vestigial nematic order above the superconducting transition temperature (though in some circumstances they
found ``joint first order superconducting transition").  In this paper, we offer
several different arguments showing that a much stronger necesssary condition is needed for
vestigial nematic order, namely $-\beta_1 < \beta_2 < -\beta_1/2$. (See Fig \ref{figp}).  Hence only systems
with parameters ``deep" in the mean-field nematic region can exhibit vestigial order. 
For  $-\beta_1 / 2 < \beta_2 < 0$ direct transition from the normal state through a second order transition into a superconducting nematic state
is expected.    Hence the experimental interpretation of \cite{Cho} of vestigial nematic order would
necessarily require a microscopic theory with parameters in that ``deep nematic region", placing
much stronger constraint on the theory themselves than the current literature realizes. 
More discussions on this will be given near the end of this paper.

The rest of the paper are as follows.
In Sec \ref{Th}, we analyze the vestigial order, using a variational method.  Besides obtaining
the condition for vestigial order mentioned just above, we also provide more details on this
vestigial transition and the superconducting transitions. 
Sec \ref{Concl} provides a conclusion.
In Appendix \ref{AppB}, we evaluate the nematic suscepibility which gives the same criterion for vestigial order
as in Sec \ref{Th}.   Appendix \ref{AppA} contains some mathematical details, as well as further discussions on parameteric dependences 
which we have left out in the text. 
  
\section{Theory for Vestigial Order}\label{Th} 

The effective Hamiltonian density 

\begin{equation}
{\mathcal H} = {\mathcal H_K} + {\mathcal H_{int}}
\end{equation}

consists of two parts.  The ``kinetic" part
\begin{widetext}
\begin{eqnarray}\label{Hk}
{\mathcal H_K} & = \alpha (\eta_j^* \eta_j) + K_1 (\partial_i \eta_j)^* (\partial_i \eta_j) + K_2  (\partial_i \eta_i)^* (\partial_j \eta_j) + K_3  (\partial_i \eta_j)^* (\partial_j \eta_i)
+ K_{zz} (\partial_z \eta_j)^* (\partial_z \eta_j)  \nonumber \\
& + \frac{K'}{2} \left[ (\partial_z \eta_y^*) (\partial_x \eta_x - \partial_y \eta_y) +    (\partial_z \eta_x^*)  (\partial_x \eta_y + \partial_y \eta_x) + h.c. \right]
\end{eqnarray}
together with the ``interacting" part
\begin{equation}\label{Hint}
{\mathcal H_{int}} = \frac{\beta_1}{2} (\eta_i^* \eta_i) (\eta_j^* \eta_j)  + \frac{\beta_2}{2} (\eta_i \eta_i)^* (\eta_j \eta_j)
\end{equation}
\end{widetext}
where sums over repeated indices $i$ or $j$ $=x,y$  are implied, and $h.c.$ denotes the hermitian conjugate. 
Effective hamiltonian density of this form has appeared in, e.g., \cite{Vanderbos16,Hecker18,Barash},  and
here we have adopted notations similar to our previous papers \cite{Nem,HQV}.
$\eta_{i}$ are the two components of the superconducting order parameter and $\partial_{i,j}$, $\partial_z$ are spatial derivatives.
$\alpha$ is a function of temperature $T$ such that $\alpha(T) > 0$ for $T > T_0$ but negative below, with $T_0$ the mean-field transition
temperture. We shall also often write $\alpha(T) = \alpha' (T - T_0)$.    
$K_{1,2,3}$ are gradient coefficients allowed in a completely cylinderically symmetric or D$_{6}$ systems.  
 The gradient term $\propto K'$ in eq (\ref{Hk}) is an addditional term allowed by the lower
D$_{\rm 3d}$ symmetry \cite{Barash,Vanderbos16,Hecker18}. 
A possible origin of this term is the fermi surface warping investigated in \cite{Akzyanov20}. 
This term affects some details of the vestigial nematic order, which we shall discuss later. 
Here we have taken the U$_2$ axis to be along $x$.
For the quartic ``interacting" terms $\beta_{1,2}$, 
in mean field, stability requires $\beta_1 > 0$, while $\beta_2 > - \beta_1$.  For $T < T_0$, the superconducting nematic state is the mean-field ground state
if $\beta_2 < 0$, whereas if $\beta_2> 0$ a superconducting state with broken time reversal symmetry would be favored. 
 Stability of the uniform state also restricts the coefficients $K_{1,2,3}$ and $K'$: we shall return to those conditions below. 

If we follow \cite{Hecker18} and introduce the column vector
\begin{equation}\label{eta}
{\bf \eta} = \left( \begin{array}{c} \eta_x \\ \eta_y \end{array} \right)  \  ,
\end{equation}
eq (\ref{Hint}) can be written as
\begin{eqnarray}
{\mathcal H_{int}} &= \frac{\beta_{1}}{2} (\eta^{\dagger} \eta)^2  +  \frac{\beta_2}{2} \sum_{\mu=x,z} (\eta^{\dagger}  \tau^{\mu}  \eta) \cdot (\eta^{\dagger} \tau^{\mu}  \eta) \label{Hinta} \\
&= \frac{\beta_{12}}{2} (\eta^{\dagger} \eta)^2  -  \frac{\beta_2}{2} (\eta^{\dagger}  \tau^y  \eta) \cdot (\eta^{\dagger}  \tau^y \eta) \label{Hintb}
\end{eqnarray}
where $\vec \tau$ are the Pauli matrices, and $\beta_{12} \equiv \beta_1 + \beta_2$.
The first line is of the same form as we have used in \cite{Shear} and the second form is the same as that in \cite{Hecker18}.
Within mean-field theory, the ground state for $\beta_2 < 0$ correspond to the column vector $\bf \eta$ being finite and real up to an overall phase factor. 
In this case, both the rotational symmetry and gauge symmetry are simultaneously broken.  The vestigial nematic phase however
correspond to the case where the expectation value of this superconducting order parameter vanishes, yet with the expectation values of 
$\eta^{\dagger} \tau^z \eta = \vert \eta_x \vert^2 - \vert \eta_y \vert^2$ and
$\eta^{\dagger} \tau^x \eta = \eta^*_x \eta_y + \eta^*_y \eta_x$  not both zero.  The finiteness of these expectation values indicate
that the rotational symmetry of the system has been broken.  \cite{Fu14,Hecker18}

In this notation,  ${\mathcal H_K}$ after Fourier transform reads
\begin{equation}\label{Hk2}
{\mathcal H_K} =  \eta^{\dagger} \left( \alpha + \epsilon_0 (\vec k) + \vec \epsilon' (\vec k)  \cdot \tau \right) \eta
\end{equation}  
$\epsilon_0 (\vec k) = \tilde K (k_x^2 + k_y^2) + K_{zz} k_z^2$ with $\tilde K \equiv K_1 + \frac{ K_{23}}{2}$,
$\vec \epsilon'$ only has $x$ and $z$ components, with $\epsilon'_z = K_{23} \frac{k^2_x - k^2_y}{2} + K' k_z k_y$,
$\epsilon'_x  = K_{23} k_x k_y + K' k_z k_x$.  $K_{23} \equiv K_2 + K_3$.  We shall assume
that, within mean-field theory, uniform states are stable even at $T_0$,  hence  for all $\vec k \ne 0$, $\epsilon_0 > 0$,
and $\epsilon_0^2 - \vec \epsilon^2 > 0$. \cite{footnote}

As we shall see, it is convenient to introduce 
\begin{equation}\label{Phi}
{\bf \Phi} = \left( \begin{array}{c} 
\Phi_{\uparrow} \\ \Phi_{\downarrow} \end{array} \right)
\equiv \frac{1}{\sqrt{2}}  \left( \begin{array}{c} 
 \eta_x + i \eta_y \\ \eta_x  - i \eta_y \end{array} \right)
\end{equation}
The fields $\Phi_{\uparrow, \downarrow}$ are, up to a factor of $\sqrt{2}$, same as the $\eta_{\pm}$ used in our earlier paper \cite{HQV}. 
The transformation from $\eta$ to $\Phi$ is similar to a different choice of quantization axis for a spin $1/2$ wavefunction,
and we shall indeed see that it is advantageous to view $\Phi$ as just forming such an object. 
\cite{transf} 
In this new basis, the kinetic part of the energy becomes
\begin{equation}\label{Hkpm}
{\mathcal H_K} =  \Phi^{\dagger} \left( \alpha + \epsilon_0 (\vec k) + \vec \epsilon (\vec k)  \cdot \sigma \right) \Phi
\end{equation}
$\vec \sigma$ are the Pauli matrices in space of (\ref{Phi}).
$\epsilon_0 (\vec k)$ is the same as before, 
$\vec \epsilon$ now only has $x$ and $y$ components, with $\epsilon_x= K_{23} \frac{k^2_x - k^2_y}{2} + K' k_z k_y$,
$\epsilon_y  =  - K_{23} k_x k_y  -  K' k_z k_x$. 
In this same basis, the interaction part of the Hamiltonian now reads
\begin{equation}\label{intpm}
{\mathcal H_{int}}  =  \frac{ g_1 }{2} (|\Phi_{\uparrow}|^4 + |\Phi_{\downarrow}|^4) + g_2 (|\Phi_{\uparrow}|^2 |\Phi_{\downarrow}|^2) 
\end{equation}
where $g_1 = \beta_1$, $g_2 = \beta_1 + 2 \beta_2$.  Now $g_1 > 0$, $g_2 > -g_1$  for stability, and the mean-field superconducting 
nematic phase is the ground state if $g_2 < g_1$.   (See Fig \ref{figp}). 
A general mean-field order parameter in this state is $\Phi$ with $\vert \Phi_{\uparrow} \vert = \vert \Phi_{\downarrow} \vert$. 
For the vestigial nematic phase, the expectation value $\langle \Phi \rangle$ of the superconducting order parameter
vanishes while $\langle \Phi^{\dagger} \sigma_{x,y} \Phi \rangle $ are not simultaneously 
zero.  Alternatively, the expectation value $\langle \Phi^*_{\uparrow} \Phi_{\downarrow} \rangle = \langle \Phi^*_{\downarrow} \Phi_{\uparrow} \rangle^*$ is finite. 
If we write $\Phi_{\uparrow(\downarrow)} = \vert \Phi_{\uparrow(\downarrow)} \vert e^{ i \chi_{\uparrow(\downarrow)}}$ 
and $\chi_{\uparrow(\downarrow)} = \chi + (-) \frac{\chi_r} {2}$ with $\chi$ an overall phase
and $\chi_r$ a relative phase, this vestigial state can be understood as one where $\chi$ is disordered whereas 
 the relative phase angle $\chi_r$ is ordered.  This state thus bears a strong similarity with the ``metallic superfluid" state studied
in, {\it e.g.} \cite{Herland10} or the ``counterflow superfluid" in \cite{Kuklov04} with here $\Phi_{\uparrow}$ and $\Phi_{\downarrow}$ playing the role of the two U(1) components there.
Now however $\chi_r$ contains information about the spatial direction in the $x-y$ plane along
which the rotational symmetry is broken (and the mechanisms considered in \cite{Herland10,Kuklov04} are also different). 

The advantage of this new basis is now obvious. As said,  one can just view the system as an effective spin-$1/2$ system.
The gradient coupling (\ref{Hkpm}) consists of 
a part $\vec \epsilon$ which can be regarded as a kind of spin-orbit coupling. 
The interaction (\ref{intpm}) in general has an XXZ symmetry.   
From eq (\ref{intpm}), it is highly suggestive that the crucial parameter which determines the 
  ``locking'' the relative phase between the $\uparrow, \downarrow$ components
is $g_2$, as we shall indeed verify below.  
This result is also supported by an examination of the ``nematic susceptibility" in App \ref{AppB}.
We remind the reader that $g_2 = 0$ ($\beta_2 = - \beta_1/2$)  corresponds to a point  ``in the middle" within the mean-field nematic region
 $-g_1 < g_2 < g_1$ ($- \beta_1 < \beta_2 < 0$).   We shall  see that it is a dividing point between where
the vestigial nematic order can exist or not.   (See Fig \ref{figp}).
In contrast, \cite{Hecker18}, employing a Hubbard-Stratanovich transformation, proposed that all regions
with $\beta_2 < 0$ can potentially exhibit vestigial nematic order.   However, the decomposition of the quartic interaction terms is not unique
(c.f. (\ref{Hinta}), (\ref{Hintb}) and (\ref{intpm})), so it is conceivable that an incorrect answer can be obtained.
We also note that, in the absence of $\vec \epsilon$ and $g_2$,   then
the up and down components are completely decoupled, and the system has an enhanced $U(1) \times U(1)$ symmetry,
where the two $U(1)$'s correspond to gauge transformations of the up and down components respectively.   In this limit
vestigial nematic order would be trivially absent. 
(For completeness, though not directly related to the problem we currently have, we mention that $g_2 = g_1$ would correspond
to a hidden $SU(2)$ symmetry; not surprising since at that point the nematic and broken time reversal symmetry states are degenerate). 

To investigate the vestigial nematic order, we employ a variational approach (see, e.g., \cite{Feynman}), which has
also been adopted before by, e.g, \cite{FB16} to study the vestigial order in the broken time-reversal symmetry case \cite{chiralnote}.
In this method, in contrast to the Hubbard-Stratanovich transformation mentioned above, 
one does not have to rely on a particular choice of writing the quartic interaction terms and an identification of
which way one is making the decomposition. 
  There exist, however,
important differences between our treatment and \cite{FB16}, on which we shall comment when we proceed. 
The free energy $F$ of a system obeys the inequality
\begin{equation}\label{var}
F \le  F_0 + \langle H - H_0 \rangle_0
\end{equation}
where $H_0$ is an ansatz Hamiltonian, $F_0$ the corresponding free energy, and the angular brackets
denote thermodynamic average performed with respect to the ansatz $H_0$, {\it i.e.}, with
the weighting factor according to $e^{ - \frac{H_0}{T}}$, where $T$ is the temperature. 
  In the notation $\Phi$,  the vestigial order
corresponds to a broken in-plane spin symmetry, hence we adopt the ansatz
\begin{equation}\label{H0}
{\mathcal H_0} = {\mathcal H_K}  - \Phi^{\dagger} (\vec h \cdot \vec \sigma) \Phi
\end{equation} 
where the in-plane vector $\vec h$ ($h_z = 0$) contains our variational parameters ($h_{x,y}$). 

The calculation can be done by noting that 
\begin{equation}\label{Gdef}
\langle \Phi_{\vec k, s} \Phi^*_{\vec k, s'} \rangle_0 \equiv T G_{s s'} (\vec k)
\end{equation}
with the ``Green's function" $\bf G$ whose inverse is given by
\begin{equation}\label{G-1}
{\bf G}^{-1} (\vec k) = \alpha + \epsilon_0 +  (\vec \epsilon - \vec h) \cdot \vec \sigma
\end{equation}
Hence
\begin{equation}\label{G}
{\bf G} (\vec k)  = \frac{ \alpha + \epsilon_0  -  (\vec \epsilon - \vec h) \cdot \vec \sigma }{ \mathcal D} \equiv G_0 + \vec G \cdot \vec \sigma
\end{equation}
with
\begin{equation}\label{D}
{\mathcal D} (\vec k)  \equiv ( \alpha + \epsilon_0)^2 -  (\vec \epsilon - \vec h) ^2
\end{equation}
$F_0$ is simply given by
\begin{equation}\label{F0}
F_0 (\vec h) = T  \ \sum_{\vec k} \ln {\mathcal D} (\vec k)  \ .
\end{equation}
Let us write 
${\mathcal H_{1}} \equiv  \frac{ g_1 }{2} (|\Phi_{\uparrow}|^4 + |\Phi_{\downarrow}|^4)$ and
${\mathcal H_{2}} \equiv  g_2 (|\Phi_{\uparrow}|^2 |\Phi_{\downarrow}|^2)$.
Now $ \langle H - H_0 \rangle_0 = \langle H_1 \rangle_0 + \langle H_2 \rangle_0 + $ $ \langle \Phi^{\dagger} \vec h \cdot \vec \sigma \Phi \rangle_0 $,
with
\begin{eqnarray}\label{H1}
\langle H_1 \rangle_0 & = g_1 \sum_{\vec k, \vec k',s = \uparrow, \downarrow} \langle \Phi^*_{\vec k, s} \Phi_{\vec k, s} \rangle_0 \langle \Phi^*_{\vec k', s} \Phi_{\vec k', s} \rangle_0 \nonumber \\
 & = 2 g_1 T^2 \sum_{\vec k} G_0 (\vec k)  \sum_{\vec k'} G_0 (\vec k')  \ ,
\end{eqnarray}
\begin{equation}\label{H2}
\langle H_2 \rangle_0 = \langle H_{2l}  \rangle_0 + \langle H_{2t}  \rangle_0 
\end{equation}
consists of a ``longitudinal" contribution
\begin{eqnarray}\label{H2l}
\langle H_{2l}  \rangle_0  &=  g_2 \sum_{k, k'} \langle \Phi^*_{\vec k, \uparrow} \Phi_{\vec k \uparrow} \rangle_0 \langle \Phi^*_{\vec k', \downarrow} \Phi_{\vec k', \downarrow} \rangle_0  \nonumber \\
& = g_2  T^2 \sum_{\vec k} G_0 (\vec k)  \sum_{\vec k'} G_0 (\vec k') 
\end{eqnarray}
and a ``transverse" piece
\begin{eqnarray}\label{H2t}
\langle H_{2t}  \rangle_0 & =  g_2 \sum_{k, k'} \langle \Phi^*_{\vec k, \uparrow} \Phi_{\vec k \downarrow} \rangle_0 \langle \Phi^*_{\vec k', \downarrow} \Phi_{\vec k', \uparrow} \rangle_0 
\nonumber \\
&=  g_2 T^2 \sum_{\vec k} \vec G (\vec k) \cdot   \sum_{\vec k'} \vec G (\vec k')   \ ,
\end{eqnarray}
and
\begin{equation}\label{h3} 
\langle \Phi^{\dagger} ( \vec h \cdot \vec \sigma) \Phi \rangle_0  =   2 T \vec h \cdot \sum_{\vec k} \vec G (\vec k)  \ .
\end{equation}
Note that our $H_{1,2}$ have been treated similarly. 

In order to see the roles of the different terms and for a closer comparison with \cite{Hecker18},  
we shall consider the various contributions to $F$ separately. 
Readers who are not interested in these details can simply note the definitions (\ref{D0}), (\ref{a}), (\ref{b}), (\ref{c}) and (\ref{c'}) below and
directly skip to eq (\ref{Final}) for the final expression for the free energy.
We  first consider only the contributions from $F_0$ (eq (\ref{F0}) ) $\langle H_{2t} \rangle$ (eq (\ref{H2t})), and
 $ \langle \Phi^{\dagger} ( \vec h \cdot \vec \sigma) \Phi \rangle_0$ (eq (\ref{h3})). 
 We expand them in $\vec h$.  For $F_0$, we get
\begin{widetext}
\begin{equation}\label{F0x}
F_0 (\vec h)  = F_0 (0) + a (h_x^2 + h_y^2) + \frac{2b}{3} (h_x^3 - 3 h_x h_y^2) + \frac{c}{2} (h_x^2 + h_y^2)^2
\end{equation} 
\end{widetext}
where we have defined
\begin{equation}\label{D0}
{\mathcal D_0} (\vec k)  \equiv ( \alpha + \epsilon_0)^2 -  \vec \epsilon^{ \ 2} \ ,
\end{equation}
\begin{equation}\label{a}
a = - T \sum_{\vec k} \left[ \frac{1}{\mathcal D}_0 + \frac { \vec \epsilon^2} { \mathcal D_0^2} \right] \ , 
\end{equation}
\begin{equation}\label{b}
b  = \frac{8}{3} T \sum_{\vec k}  \frac {  \epsilon_x^3} { \mathcal D_0^3}  \ , 
\end{equation}
and
\begin{equation}\label{c}
c = - T \sum_{\vec k} \left[ \frac{1}{\mathcal D_0^2 } +  4  \frac {\vec  \epsilon^2} { \mathcal D_0^3}  +  3  \frac { \vec \epsilon^4} { \mathcal D_0^4}\right] \ .
\end{equation}
In obtaining eq (\ref{F0x}), we have made use of the D$_{3d}$ symmetry of the crystal to relate some of the sums (See App \ref{AppA}). 
We also remark that $b$ is non-zero only when $K'$ is finite (see also App \ref{AppA}). 
We note that eq (\ref{F0x}) obeys  D$_{3d}$ symmetry, in particular, $(h_x^3 - 3 h_x h_y^2)$ is an allowed cubic invariant,
as it remains the same under rotation by $2 \pi/3$ about the $z$ axis and rotation by $\pi$ about $x$. 

In eqs (\ref{H2t}) and (\ref{h3}), we need also the sums $T \sum_{\vec k} G_{x,y} (\vec k) $.    On noting that
$d {\mathcal D} / d h_x = 2 (\epsilon_x - h_x)$ and recalling eq (\ref{G}), we see that they can be obtained
simply by differentiating $T  \sum_{\vec k} \ln {\mathcal D} (\vec k)$ hence eq (\ref{F0x}) with respect to $h_{x,y}$ and then multiplying by $-1/2$. 
We get eventually
\begin{widetext}
\begin{equation}\label{H2tf}
\langle H_{2t}  \rangle_0 = g_2 \left[  a^2 (h_x^2 + h_y^2) + 2 a b  (h_x^3 - 3 h_x h_y^2) +  (2 a c + b^2)  (h_x^2 + h_y^2)^2 \right]
\end{equation}
and 
\begin{equation}\label{h3f}
\langle \Phi^{\dagger} ( \vec h \cdot \vec \sigma) \Phi \rangle_0 = 
 - 2 a (h_x^2 + h_y^2)  - 2  b  (h_x^3 - 3 h_x h_y^2) - 2 c (h_x^2 + h_y^2)^2 \ .
\end{equation}
These three contributions $F_0$, $\langle H_{2t} \rangle$, and
 $ \langle \Phi^{\dagger} ( \vec h \cdot \vec \sigma) \Phi \rangle_0$  together give an interim free energy,  which we shall call $F_{interim}$,  
\begin{equation}\label{Finterim}
F_{interim} (\vec h) =  F_{interim} (0)  + [ a (g_2 a - 1)] (h_x^2 + h_y^2)  + [  b  ( 2 g_2 a - \frac{4}{3})] (h_x^3 - 3 h_x h_y^2) + 
[g_2 (2 a c + b^2) - \frac{3 c}{2} ] (h_x^2 + h_y^2)^2  \ .
\end{equation}
\end{widetext}

Let us analyze $F_{interim}$ and pretend this is the full expression for $F$ at the moment. 
Let us first note that, for temperatures above the mean field transition temperature $T_0$, $\alpha > 0$ and hence 
$a$ is negative definite.   We see that if $g_2 > 0$, the coefficient of the $\vec h^2$ term is positive definite.  
$h=0$ is always a local minimum and no broken symmetry state with finite $\vec h$ is expected for 
temperatures above $T_0$. 
If $g_2 < 0$, the situation is different. Writing it as $(-a) \vert g_2 \vert ( \frac{1}{\vert g_2 \vert } + a)$, noting that since the magnitude of $a$ increases
as the temperature is lowered towards $T_0$ (and diverges to $-\infty$ at $T_0$ where $\alpha \to 0$), we see that this coefficient is
 positive at high temperatures, then vanishes at a ``critical temperature" $T_1 > T_0$
where $ \tilde a \equiv ( \frac{1}{\vert g_2 \vert } + a) = 0$, and changes sign  below.  This indicates 
a possible broken symmetry state above the mean-field transition temperature $T_0$. 
$g_2 < 0$ is required, in agreement with App \ref{AppB}. 

Let us, in the spirit of Ginzburg-Landau theory, approximate all coefficients by the value at $T_1$ except the coefficient of $\vec h^2$,
that is, in all terms except $\tilde a$, put $g_2 a = 1$.  We get
\begin{widetext}
\begin{equation}\label{Finterimf}
F_{interim} (\vec h) \approx   F_{interim} (0) +   ( \frac{1}{\vert g_2 \vert } + a) (h_x^2 + h_y^2)  +  \frac{2}{3} b (h_x^3 - 3 h_x h_y^2) + 
[g_2 b^2 + \frac{ c}{2} ] (h_x^2 + h_y^2)^2
\end{equation}
\end{widetext}

At this point, it is interesting to compare this result with what we would get if we treat the $g_2$  interaction term 
by  a Hubbard-Stratanovich transformation (ignoring ${\mathcal H_1}$ for the moment).
If we write ${\mathcal H}_2$ as $ g_2 ( \Phi^*_{\uparrow} \Phi_{\downarrow})  ( \Phi^*_{\downarrow} \Phi_{\uparrow})
= \frac{g_2}{4} \sum_{\mu = x,y} ( \Phi^\dagger \sigma^{\mu} \Phi)  ( \Phi^\dagger \sigma^{\mu} \Phi)$
and decompose this quartic term using $\frac{\vec h^2}{(-g_2)} - \vec h \cdot  ( \Phi^\dagger \vec \sigma \Phi)$
with $\vec h$ containing again only $x$ and $y$ components, we obtain an effective Hamiltonian
\begin{equation}\label{Heff}
{\mathcal H}_{eff} = {\mathcal H_K} - \vec h \cdot  ( \Phi^\dagger \vec \sigma \Phi) +  \frac{\vec h^2}{(-g_2)} 
\end{equation}
 Now given $H_{eff}$ and $g_2 < 0$, the free energy
is simply 
\begin{equation}\label{Feff}
F_{eff} = \frac{ \vec h^2}{\vert g_2 \vert} + T \sum_{\vec k} \ln {\mathcal D} (\vec k)
\end{equation}
If we expand this expressions in $\vec h$ (noting that the last term is just the same as our $F_0$ in eq (\ref{F0}) and hence eq (\ref{F0x})),
we obtain an expression that is identical with eq (\ref{Finterimf}) except that the $g_2 b^2$ term (which is typically small and is absent
entirely if the symmetry is slightly higher, say $D_{6h}$, see App. \ref{AppA})  in front of $\vec h^4$ is now absent. 
We can trace the reason for this similarity by noting that, if we put $g_2 a = 1$, the sum of 
$H_{2t}$ and $ \langle \Phi^{\dagger} ( \vec h \cdot \vec \sigma) \Phi \rangle_0$  is just (see eqs (\ref{H2tf}) and (\ref{h3f}))
$ \frac{ \vec h^2}{\vert g_2 \vert}$, apart from the $g_2 b^2 \vec h^4$ term we just mentioned (there are further differences but higher orders in $\vec h$). 
Taking the derivative of eq (\ref{Feff}) we obtain
a self-consistent equation for $\vec h$, which reads
\begin{equation}\label{self}
\vec h = g_2 \ T \sum_{\vec k} \frac{ \vec \epsilon - \vec h}{ \mathcal D}
\end{equation} 
This has the same form as the self-consistent equation in \cite{Hecker18}, except the important difference that the interaction
coefficient appearing here is $g_2$ , while the expression in \cite{Hecker18} contains what is  $\beta_2$ in our notation.
 This difference is an artefact of the Hubbard-Stratanovich decoupling procedure mentioned earlier:
 the decomposition of the quartic term depends spuriously on the way one chooses to express the term.
(Also, on  the right-hand-side of eq (\ref{self}), 
instead of our  ${\mathcal D}$ in the denominator, they have instead
$(\alpha + R + \epsilon_0)^2 - \vec \epsilon^{\ 2}$, thus with an extra contribution $R$.  We shall 
comment on this difference later). 
We note that eq (\ref{self}) implies the same condition for vestigial nematic order as we found earlier: $g_2 < 0$, or $\beta_2 < -\beta_1/2$, rather than just $\beta_2 < 0$ 
found in \cite{Hecker18}.

However, there is a serious problem in this simplified analysis so far.  
While eq (\ref{Finterim})-(\ref{self})  seemingly yield the correct condition for vestigial order, 
we will shortly see that $F_{interim}$ in eq (\ref{Finterimf}) does not have a stable ground state. 
 (As corollary, any theory based solely on eqs (\ref{Feff}) and (\ref{self}) must also be unstable.)
We shall see that the terms $\langle H_{1} \rangle$ and $\langle H_{2l} \rangle$  that we have left out thus far, stabilize the theory.
Note then that since $F_{interim}$ in eq (\ref{Finterimf}) is not our full expression for
the free energy $F$, and since $\vec h$ should be determined from the minimization of $F$, eq (\ref{self}) is {\em not}
our equation for $\vec h$.   However, as we shall see shortly below, the $\vec h^2$ coefficient of $F$ {\em is}
correctly given by that in eq (\ref{Finterim}) thus (\ref{Finterimf}),  
hence it does not alter the fact that $g_2 < 0$ is needed for vestigial nematic order. 

Let us return to $F_{interim}$ in  eq (\ref{Finterimf}).  We see that there is a serious problem: the coefficient of the fourth order term
is negative (see (\ref{b}) and (\ref{c})).  In fact, one can show that the coefficients of all $\vec h^{2N}$ terms with $N \ge 2$ are negative.
We now show that  the contributions $\langle H_{1} \rangle$ and $\langle H_{2l} \rangle$
we have left out stabilize the theory.   Expanding them in $\vec h$, we obtain 
\begin{widetext}
\begin{equation}\label{expand}
 \langle H_1 + H_{2l} \rangle = ( 2 g_1 + g_2 ) T^2 \left( \sum_{k} (G_0 \vert_{\vec h=0} + \vec h \cdot \frac{ \partial  G_0}{ \partial \vec h} + ...) \right) \times 
\left( \sum_{k'} (G_0\vert_{\vec h=0} + \vec h \cdot \frac{ \partial G_0}{ \partial \vec h} + ...) \right)
\end{equation} 
\end{widetext}
with the $\vec h$ derivatives evaluated at $\vec h = 0$.  
Since $G_0 \propto 1/\vec k^2$ at large $\vec k$, we see that the sum
$T \sum_{\vec k} G_0 \vert_{\vec h = 0}$ is ultraviolet divergent. 
The $\vec h = 0$ contribution is however irrelevant to us since we only need to consider $F (\vec h) - F(\vec h =0)$.
There is no first order term in eq (\ref{expand}) as
$ \frac{ \partial  G_0}{ \partial \vec h} \vert_{\vec h=0} = - 2 \frac{ \alpha + \epsilon_0}{\mathcal D_0^2} \vec \epsilon (\vec k)$
sums to zero due to the angular dependent $\vec \epsilon$.   At first sight one might think there is an $\vec h^2$ contribution from 
$T^2 \left( \sum_{k} G_0 \vert_{\vec h=0} \right)  \times  \left( \sum_{\vec k'}   \frac{ h_{\mu} h_{\nu}}{2} \frac{ \partial^2  G_0}{ \partial h_{\mu} \partial h_{\nu}} \right)$  or vice versa.
However, one can easily see that these terms are just what we would get for the modifications to the $\vec h^2$ terms of $F_0$ if we include the one-loop self energy terms due to
$g_{1,2}$ in  ${\bf G}$, {\it i.e.}, if we insert a self-energy
 $ - 2 g_1 \langle \Phi_{\uparrow} (\vec k') \Phi^*_{\uparrow} (\vec k') \rangle_{\vec h=0}  $ $-  g_2  \langle \Phi_{\downarrow} (\vec k') \Phi^*_{\downarrow} (\vec k') \rangle_{\vec h=0} $ 
in the $\uparrow \uparrow$ component ${\bf G^{-1}}$ of eq (\ref{G-1}) (and similarly for $\uparrow \leftrightarrow \downarrow$).
Including this self-energy is equivalent to replacing $\alpha$ by $\alpha + ( 2 g_1 + g_2) T \sum_{\vec k} G_0 ({\vec k}) \vert_{\vec h=0}$. 
These insertions simply renormalizes $T_0$ and $\alpha'$, that is, the mean-field transition temperature and the derivative of $\alpha$ with respect to the temperature.
As in usual treatment of phase transitions \cite{PP,Z}, we assume that these replacements have already done from the outset and therefore we shall simply 
leave this contribution out.  There are therefore no modifications to $F(\vec h)-F(0)$ that is second order in $\vec h$.

Neither there are modifications to $F(\vec h)-F(0)$ of third order since $\sum_{\vec k}  \frac{ \partial  G_0}{ \partial \vec h} \vert_{\vec h=0}$ vanishes
as explained above.  The lowest order contribution is thus fourth order in $\vec h$, arising from
\begin{widetext}
$ T^2 \left( \sum_{k}  \frac{ h_{\mu} h_{\nu}}{2} \frac{ \partial^2  G_0}{ \partial h_{\mu} \partial h_{\nu}}  \right) 
\times \left( \sum_{k}  \frac{ h_{\mu'} h_{\nu'}}{2} \frac{ \partial^2  G_0}{ \partial h_{\mu'} \partial h_{\nu'}} \right)$.
The factor $ T^2 \left( \sum_{k}  \frac{ h_{\mu} h_{\nu}}{2} \frac{ \partial^2  G_0}{ \partial h_{\mu} \partial h_{\nu}}  \right)$ is finite only for $\mu=\nu=x$ or $y$, and we get the contribution
$\frac{c'}{2} \vec h^4$ with \cite{g1h}
\begin{equation}\label{c'}
c' = 2 ( 2 g_1 + g_2) T^2 \left[  \sum_{\vec k} \left( \frac{ \alpha + \epsilon_0}{\mathcal D_0^2} + 2  \frac{ \alpha + \epsilon_0}{\mathcal D_0^3} \vec  \epsilon^2 \right) \right]^2
\end{equation}  

  The end result is that  the free energy $F$ is given by eq (\ref{Finterimf}) with an additional contribution to the fourth order term, thus
\begin{equation}\label{Final}
F (\vec h) =  F(0) +  \tilde a (h_x^2 + h_y^2)  +  \frac{2}{3} b (h_x^3 - 3 h_x h_y^2) + 
\frac{ \tilde c}{2}  (h_x^2 + h_y^2)^2
\end{equation}
\end{widetext}
where $\tilde c = c + 2 g_2 b^2 + c'$.  We remind the readers that 
$\tilde a \equiv ( \frac{1}{\vert g_2 \vert} + a ) $ is positive for $T > T_1$ and negative below, with $T_1 > T_0$. 

We note here all the coefficients $\tilde a$, $b$, $\tilde c$ entering eq (\ref{Final}) are given by sums that
are ultraviolet convergent: to compute them, one needs only the information near $\vec k \approx 0$. 
  This is in contrast to both \cite{Hecker18} and \cite{FB16}.  They both have explicitly 
included a term that correspond to our one-loop self-energy mentioned in the discussion below (\ref{expand}).
This term has been removed by us by renormalization of $\alpha$.  The treatment of this term in this
way is also consistent with App. \ref{AppB}.

The stability $\tilde c > 0$  is provided by $g_1 > 0$ if $g_1$ is sufficiently large.   Let us examine this condition in more detail.
The presence of $\vec \epsilon$ in ${\mathcal D_0}$ makes the analytic valuation of the integrals difficult.
Let us first simplify the problem first by pretending that the $\vec \epsilon^2$ term in ${\mathcal D_0}$ is small,
and replace all ${\mathcal D_0}$ terms in the denominators of the sums involved by 
 ${\mathcal D_{00}} \equiv (\alpha + \epsilon_0)^2$.   We find (see App \ref{AppA})
\begin{equation}\label{c1}
c = - \frac{1}{2^6 \pi}  \frac{ T} { \alpha^{5/2} \  \tilde K  \ K_{zz}^{1/2}} 
\end{equation}
and
\begin{equation}\label{c'1}
c' = 2 (2 g_1 + g_2) \left[  \frac{1} {2^5 \pi}  \frac{ T} { \alpha^{3/2} \  \tilde K  \ K_{zz}^{1/2}} \right]^2
\end{equation}
In these expressions, we have only kept the first terms in eq (\ref{c}) and (\ref{c'}), ignoring the terms involving explicitly $\vec \epsilon$'s in
the same spirit as just described.   Note that then, as the temperature is lowered towards the mean-field transition temperature from above,
the magnitude of  $c'$ grows faster than $c$. 
On the other hand,
 the temperature $T_1$ where the term $\tilde a$ in $F$
changes sign occurs at (see App \ref{AppA})
\begin{equation}\label{T1}
\frac{1}{\vert g_2 \vert} = \frac{1}{8 \pi}  \ \frac{ T_1} { \alpha(T_1)^{1/2} \  \tilde K  \ K_{zz}^{1/2}}  \ , 
\end{equation} 
and hence
\begin{equation}\label{T10}
(T_1 - T_0) = \frac{1}{\alpha'} \frac{ ( \vert g_2 \vert T_1 )^2 } { \tilde K^2 K_{zz} }  \frac{1}{(8 \pi)^2} \ .
\end{equation}
where on the right hand side we can also replace the explicit temperature $T_1$ by $T_0$ since the dominant temperature
variation in eq (\ref{T1}) arises from $\alpha(T)$. One recognizes the right hand side has the same parametric form of the usual (Ginzburg) estimate for the width of the fluctuctation region \cite{LL}
with $g_2$ playing the role of the interaction.   
For usual superconductors this region is expected to be small compared with the mean-field transition temperature $T_0$, though \cite{Hecker18} 
obtained a rather large value in their theory of doped Bi$_2$Se$_3$.
If we replace the coefficients $c$ and $c'$ by their values at $T_1$ (in the spirit of usual Ginzburg-Landau theory), the condition $\tilde c > 0$ is equivalent to
(dropping the contribution $g_2 b^2$ in the same spirit as above)
$ g_1 >  \vert g_2 \vert$, hence satisfied for the entire region where the mean-field theory is stable.
If we include the contributions from $\vec \epsilon$,  $\tilde c > 0$ will continue to hold except perhaps for some violation near $g_2 \approx - g_1$.

Assuming $\tilde c > 0$, the 
  the analysis of  the free energy (\ref{Final})  is standard.
In the special case  $b = 0$, (recall this is the case if $K'=0$) then we have a second order transition into the vestigial nematic state with $\vec h \ne 0$ at $T_1$, where  $\tilde a$ changes sign.
For the more general situtation with $b \ne 0$, we instead obtain a first order phase transition from the normal state to the vestigial nematic state
at $\tilde a(T_1^*) = \frac{2}{9} \frac{b^2}{\tilde c} > 0$, hence $T_1^* > T_1$, to the state $\vec h = h_x \hat x$ (or its rotated partners by $\pm 2 \pi/3$)
with $h_x = - \frac{2}{3} \frac{b}{\tilde c}$.   
$b$ is finite only when both $K_{23}$ and $K'$ are finite, but is even in $K'$ while odd in $K_{23}$, with
${\rm sgn} b = - {\rm sgn} (K_{23})$ (see App \ref{AppA}), hence  ${\rm sgn} h_x =  {\rm sgn} (K_{23})$ \cite{reflect} .
To be self-consistent, the above assumed that the value of $\vert \vec h \vert =  \vert \frac{2 b} { 3 \tilde c}\vert $ at $T_1^*$ is less than $\alpha(T_1^*)$, so that 
${\mathcal D}(\vec k) $ at this point is still positive, else we should have a first order phase transition directly into a superconducting state
with broken rotational and broken gauge symmetry.  For more discussions on this condition, see App \ref{AppA}. 

Upon lowering the temperature from $T_1^*$, $\alpha(T)$ decreases but $\vert \vec h \vert$ increases, 
hence at some temperature $T_c^* < T_1^*$, ${\bf G}^{-1}(\vec k) $ will have a zero eigenvalue.  
$|{\bf \Phi}|$ grows from $0$ at $T_c^*$ and increases with lowering temperature, 
signalling a second order transition into the superconducting state.
This transition turns out
to occur at $\vec k = 0$ and at the temperature $T_c^*$ where  $\alpha (T_c^*) = \vert \vec h \vert > 0$ .  
To check this, consider the special case $\vec h = h_x \hat x$. 
Then ${\mathcal D}(\vec k=0)  = \alpha^2 - h_x^2$, thus vanishes at $\alpha = h_x$.
For general $\vec k$,  ${\mathcal D}(\vec k) = \alpha^2 - h_x^2 + 2 (\alpha \epsilon_0 + h_x \epsilon_x) + (\epsilon_0^2 - \vec \epsilon ^2 )$.
If $\vec k \ne 0$, 
the last term is positive by our assumption.  At $ \vert h_x \vert = \alpha$, the second term is also positive due to the same criterion.
Hence at $\alpha = h_x$,  ${\mathcal D}(\vec k)  > 0$ if $\vec k \ne 0$, hence the transition occurs at $\vec k=0$ as claimed.

\begin{figure*}
\includegraphics[width=0.7\textwidth]{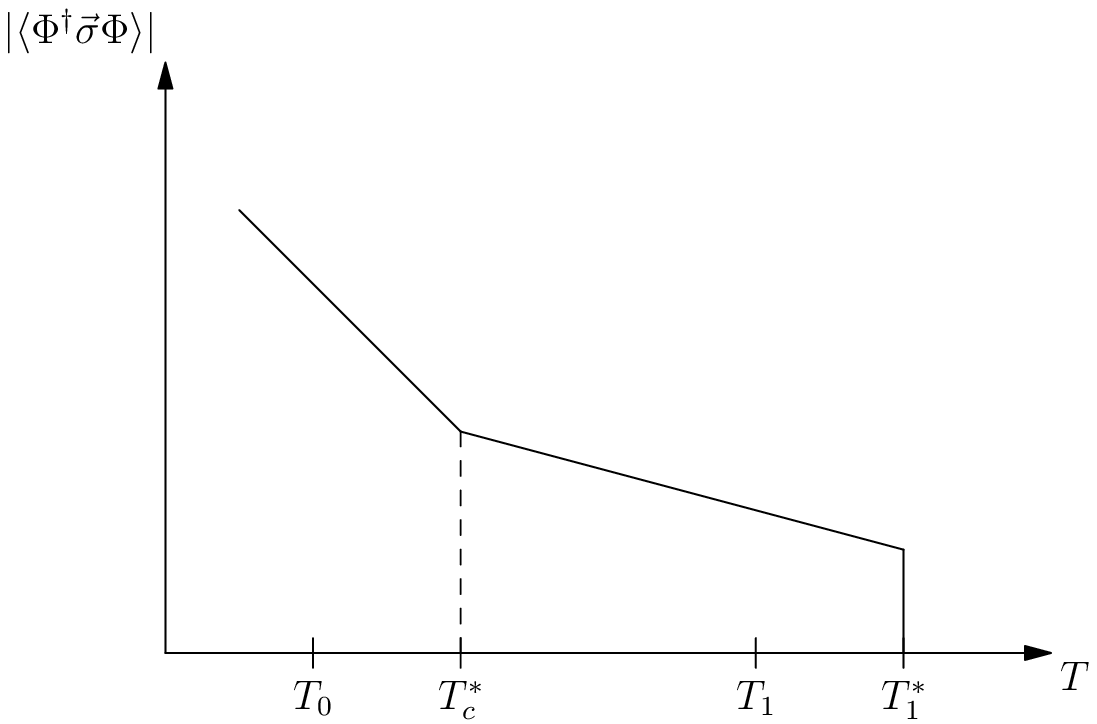}
\caption{
 Schematic behavior of $\vert \langle \Phi^{\dagger} \vec \sigma \Phi \rangle\vert $ as a function of temperature in the case where the conditions discussed in text are fulfilled.
(one necessary condition being $g_2 < 0$).
$T_0$: mean-field transition temperature of the superconductor, {\it i.e.}, where $\alpha(T)$ changes sign. 
$T_1$ is where the coefficient $\tilde a(T)$ changes sign.  This would be the phase transition to the vestigial nematic order if the transition were second order.
$T_1^*$ is the first order phase transition temperature into the vestigial nematic state.  $T^*_{c}$ is the transition temperature to the superconducting state from
the vestigial nematic state. }
\label{fig1}
\end{figure*}

At the transition, the superconducting state $\Phi_{\vec k=0}$ is an eigenvector of  ${\bf G}^{-1}(\vec k=0) $ with a zero eigenvalue.
Since at this point  ${\bf G}^{-1}(\vec k=0)  = \alpha - h_x \sigma^x$,  we have $h_x \sigma^x {\bf \Phi}_{\vec k=0} = \alpha {\bf \Phi}_{\vec k=0}$
Hence $\Phi^{\dagger}_{\vec k=0} \sigma^x \Phi_{\vec k=0}$ has the same sign as $h_x$  thus also
the expectation value $\sum_{\vec k} \langle \Phi^{\dagger}_{\vec k} \sigma^x \Phi_{\vec k} \rangle$  from the finite $\vec k$ modes.
We sketch the expected behavior in Fig \ref{fig1}.   We have not yet developed a theory for $T < T_c$. 

\section{Conclusions}\label{Concl}

In this paper, we examine carefully the condition of vestigial nematic order for a nematic superconductor.
While the nematic superconducting ground state is expected for $-\beta_1 < \beta_2 < 0$, only
the ``deeper" part of this region with $-\beta_1 < \beta_2 < -  \beta_1/2$ ($g_2 < 0$) can exhibit vestigial nematic order
above the superconducting state.   The interpretation of the experiment \cite{Cho}, if correct, would
exclude a large region of parameter space.   Conversely, if the microscopic theory can constraint these
parameters to the alternate region, then a different interpretation of the results in \cite{Cho} must be sought. 
$\beta_{1,2}$ in particular depend on the momentum and spin structure of the order parameter, and
many model calculations have been given in the literature  \cite{Zyuzin,Yuan,Uematsu,VanderbosR}.
In \cite{Zyuzin}, two models are studied, but both of them have $g_2 > 0$.   
 Refs. \cite{Yuan,Uematsu,VanderbosR} plotted phase diagrams containing both  nematic and chiral phases,
but they did not indicate explicitly the positions corresponding to $g_2 = 0$.   However, 
since large regions of their nematic phases actually border the chiral phase,  we know at least 
 that those regions cannot exhibit vestigial nematic order. \cite{noteVR} 

We remark that this is not the only example where a nematic superconductor behaves qualitatively differently according to
the parameters $\beta$'s.  Previously, when investigating the stability of half-quantum vortices near
the lower critical field \cite{HQV}, we found that they are always stable for $g_2 > 0$.  On the other hand,  two half-quantum vortices might ``collapse" back to an ordinary phase vortex
if $g_2 < 0$, unless counter-balanced by sufficiently large $K_{23}$. 
Thus to understand the properties of a nematic superconductor and thus doped Bi$_2$Se$_3$, it is crucial to discern in which
parameter region the system lies, and whether and how this depends on parameters such as doping concentrations. 

\section{Acknowledgements}

This work is supported by the Ministry
of Science and Technology, Taiwan under Grant No.  MOST-110-2112-M-001-051 -MY3, and P.T.H. is supported under
Grant No. MOST 110-2811-M-001-561.

\appendix

\section{Nematic Susceptibility} \label{AppB}

Here we want to verify  the condition $g_2 < 0$ for vestigial nematic order by evaluating the ``nematic susceptibility", 
in particular we would like to check that this is not an artifact of the particular basis we have chosen. 
We thus now use the original $\eta$ notation, thus Hamiltonian eq (\ref{Hinta}) and (\ref{Hk2}).  
We evaluate the susceptibility to  an external field coupling to $ \eta^{\dagger} \tau_x \eta$ with zero external momentum.  This susceptibility, in the
random phase approximation, is given by the product of two Green's function with an external vertex $\tau_x$ and
a renormalized vertex $\Gamma^{(x)}$, which is given by the Bethe-Salpeter equation
\begin{widetext}
\begin{eqnarray}\label{Gamma}
\Gamma^{(x)}_{ij}  & = (\tau^x)_{ij}  -  \beta_1 \delta_{ij} T \sum_{\vec k'} \left[ G_{l_1 l}(\vec k') \Gamma^{(x)}_{l l'} G_{l' l_1} (\vec k') \right]  
 -  \beta_1 T \sum_{\vec k'} \left[ G_{il}(\vec k') \Gamma^{(x)}_{l l'} G_{l'j} (\vec k') \right]  \nonumber \\
&  - \beta_2 T \sum_{\vec k'} \tau^{\mu}_{ij}   \left[  \tau^{\mu}_{l l'} G(\vec k')_{l' l_1}  \Gamma^x_{l_1 l_2}  G_{l_2 l} (\vec k') \right]
   -\beta_2 T \sum_{\vec k'} \left[ \tau^{\mu}_{ii'} G_{i'l}(\vec k')  \Gamma^{(x)}_{l l'} G_{l' j'} (\vec k')  \tau^{\mu}_{j' j} \right]
\end{eqnarray}
\end{widetext}
where $i,j$ runs over the two components in $\eta$ space and $\mu=x,z$, here ${\bf G}$ is the Green's function for $\eta$ in zero field,
that is $ {\bf G} (\vec k)  = \frac{ \alpha + \epsilon_0  -  \vec \epsilon'\cdot \tau }{ \mathcal D_0}$ ({\it c.f.} eq (\ref{Hk2}))
One can check that ${\bf \Gamma}^x$ is proportional to $\tau^x$, so let us denote this coefficient by  $\Gamma^{x(x)}$.
It is convenient to write ${\bf G} = G_0 + G_x \tau^x + G_z \tau^z$. With this, we see that the first interacting term does not contribute,
and the last term in eq (\ref{Gamma}) vanishes after sum over $\mu$, and we obtain the
self-consistent equation
\begin{equation}
\Gamma^{x(x)} =  1  - ( \beta_1+2 \beta_2)  T \sum_{\vec k'} \left[ G_0 G_0 -G_z G_z + G_x  G_x  \right] \Gamma^{x(x)}
\end{equation}
where we have left out the arguments $(\vec k')$ of $G_0$ etc for simplicity. 
Since the sums $  \sum_{\vec k'} (G_x G_x) $ and $  \sum_{\vec k'} (G_z G_z) $ are equal, we get 
\begin{equation}
\Gamma^{x(x)} = \left[ 1 + (\beta_1 + 2 \beta_2) T  \sum_{\vec k'} ( G_0 G_0)  \right]^{-1}
\end{equation}
Hence the vertex $\Gamma^{x(x)}$ and the susceptibility diverges at
\begin{equation}\label{div}
1 + g_2 T \sum_{\vec k} \frac{ (\alpha + \epsilon_0)^2} { \mathcal D_0^2} = 0
\end{equation}
This is possible only if $g_2 < 0$, and in that case, eq (\ref{div}) 
is the same condition as $\tilde a = 0$.  Note that, using (\ref{D0})
$a$ of eq (\ref{a}) can also be rewritten as $ a = - T  \sum_{\vec k} \frac{ (\alpha + \epsilon_0)^2} { \mathcal D_0^2}$.

We obtain exactly the same criterion if we consider the response to $\tau^z$. 
In the above we have evaluated the nematic suceptility for an non-interacting system.  If we insert one-loop
self-energies to the propagators, we would only modify the $\alpha$'s in ${\bf G}(\vec k)$ to 
 $\alpha + ( 2 g_1 + g_2) T \sum_{\vec k} G_0 ({\vec k})$.  This just replaces these $\alpha$'s by the
effective ones  and thus does not affect the requirement that $g_2$ has to be negative for the divergence
of the nematic succeptibility. 

\section{Mathematical Details and Further Estimates} \label{AppA}

We first consider some symmetry properties.   Under a $2 \pi/3$ rotation,
we map $(k_x, k_y)$ to $(k'_x, k'_y) = (c k_x -s k_y, s k_x+ c k_y)$ with $c \equiv \cos ( 2 \pi/3)$ and 
$s \equiv \sin (2 \pi/3)$.  Correspondingly $k_x \pm i k_y \to ( k_x \pm i k_y) \omega^{\pm 1}$ where
$\omega \equiv e^{ 2 \pi i / 3}$.  Since $\Phi_{\uparrow, \downarrow} \propto ( \eta_x \pm i \eta_y)$,
we have $\Phi_{\uparrow,\downarrow} \to \Phi_{\uparrow,\downarrow} \omega^{ \pm 1}$.
Also $G_{\uparrow, \downarrow} \to \omega^{-1} G_{\uparrow, \downarrow} $,
corresponding $ (h_x \pm i h_y) \to (h_x \pm i h_y) \omega^{\pm 1}$. 

The symmetry property of $\epsilon_x \pm i \epsilon_y$ follows from that of $ ( k_x \pm i k_y) $
(note the negative sign in the definition of $\epsilon_y$): 
$\epsilon_x \pm i \epsilon_y \to ( \epsilon_x \pm i \epsilon_y ) \omega^{\pm 1}$.  Hence
$\vec \epsilon$ transform in the same manner as $\vec h$, with the two components 
transforming as $(k_x^2 - k_y^2, - 2 k_x k_y)$ under  D$_{3d}$.  

For the momentum sums, we note that ${\mathcal D_0}$ is an invariant. 
It follows immediately that sums of the form 
$\sum_{\vec k} \frac{ (\epsilon_x \pm i \epsilon_y)^j}{\mathcal D_0^n}$ vanish unless
$j$ is a multiple of $3$.   From these we see that
$\sum_{\vec k} \frac{ \epsilon_x}{\mathcal D_0^n} = \sum_{\vec k} \frac{  \epsilon_y}{\mathcal D_0^n}= 0$,
whereas 
$\sum_{\vec k} \frac{ (\epsilon_x)^2}{\mathcal D_0^n}= \sum_{\vec k} \frac{  (\epsilon_y)^2}{\mathcal D_0^n} $,
Also, using the transformation property of $\epsilon_{x,y}$, we obtain
$\sum_{\vec k} \frac{ (\epsilon_x)^3}{\mathcal D_0^n} = - \sum_{\vec k} \frac{ \epsilon_x (\epsilon_y)^2}{\mathcal D_0^n}$,
and 
 $\sum_{\vec k} \frac{ (\epsilon_x)^4}{\mathcal D_0^n}= \sum_{\vec k} \frac{  (\epsilon_y)^4}{\mathcal D_0^n}
= 3 \sum_{\vec k} \frac{ (\epsilon_x \epsilon_y)^2}{\mathcal D_0^n} $.

We now turn to the evaluation of some of the sums and integrals.

Let us consider $a$ of eq (\ref{a}), and approximate the denormator ${\mathcal D_0}$ there by ${\mathcal D_{00}}$ as discussed in text.
That is, we would like to calculate the sum
$ - T \sum_{\vec k} \left[ \frac{1}{\mathcal D_{00}} \right]$. To do this, we introduce
$x = \left( \frac{\tilde K}{\alpha} \right)^{1/2} k_x$ and similarly for $x \to y$, and $z = \left( \frac{K_{zz} }{\alpha} \right)^{1/2} k_z$.
This sum then becomes
\begin{widetext}
\begin{equation}\label{a-cal}
 -  T  \frac{  \alpha^{3/2}} { \tilde K  \ K_{zz}^{1/2}} \frac{1}{\alpha^2}  \int  \frac{d^3 x}{(2 \pi)^3} \frac{1}{ ( 1 + R^2)^2} 
\end{equation}
where $R^2 \equiv (x^2 + y^2 + z^2)$.  The integral gives $ 1/ (8 \pi)$ hence eq (\ref{T1}). 
The terms in eq (\ref{c1}) and (\ref{c'1}) are obtained in similar manner.

Let us examine the second contribution to $a$ in eq (\ref{a}). That is, $ - T \sum_{\vec k}  \frac { \vec \epsilon^2} { \mathcal D_0^2} $.
Similar to above, we first replace the denormator ${\mathcal D_0}$ there by ${\mathcal D_{00}}$ and use the same substitutions as above.
After this we can replace $(x^2 - y^2)^2$ etc by their angular averages.  We obtain the contribution
\begin{equation}\label{a2}
 -   T  \frac{ \alpha^{3/2}} { \tilde K  \ K_{zz} ^{1/2}} \frac{1}{\alpha^2} \left[ \left(\frac{K_{23}}{\tilde K} \right)^2 + \frac{( K')^2}{K_{zz} \tilde K} \right] 
  \frac{2}{15}  \int   \frac{d^3 x}{(2 \pi)^3} \frac{R^4}{ ( 1 + R^2)^4}
\end{equation}
\end{widetext}
The factor $2/15$ is from the angular average.  Note that the last term has the  same large $R$ dependence as eq (\ref{a-cal}) but has higher $R$ powers at $R \to 0$.
The integral gives $5 / (2^6 \pi)$.  Hence this contribution is much smaller than the one given in (\ref{a-cal}) even when
the quantity in the square bracket of eq (\ref{a2})
is of order $1$.
(the correction is  $ \frac{2}{15} \times \frac{5}{2^6 \pi} \times 8 \pi =  \frac{1}{12}$ of the original) . 
This is because of (i) the angular average and (ii) the smaller $d^3 x$ integral, which is in turn due to the higher powers in $R$
arising from the $\epsilon_x^2$ factor. 

Similar remarks apply to the other terms in, e.g., eq (\ref{c}) and (\ref{c'}). 
Note also that, when we restore the $\vec \epsilon^2$ in the denomintors  $\mathcal D_0$ but expand in it, the correction terms
are exactly of the same forms as the ``higher order" terms in these equations. 
 Hence we conclude that, unless in extreme circumstances
of very large $K_{23}$ compared with $\tilde K$ etc, the condition for $\tilde c > 0$ is, to a good approximation, given 
as in text.  (Gradient terms were evaluated in, e.g., \cite{Zyuzin} for two models, giving
$K_{23}/\tilde K = 1$ and $2/3$; $K'$ was not given there)
 Note also that the condition $\epsilon_0^2 > \vec \epsilon^2$ limit the sizes of  
$K_{23}/\tilde K$ and $K'{}^2 / (\tilde K K_{zz})$.   That is, unless the system is close to one where the net gradient
energy is small along some momentum directions, the stability condition we gave is a good approximation. 

Now let us turn to $b$ in eq (\ref{b}).  Replacing $\vec k$ by $x,y,z$ as explained above, the sum
$ \sum_{\vec k}  \frac {  \epsilon_x^3} { \mathcal D_0^3} $ becomes
\begin{widetext}
\begin{displaymath}
 \frac{1}{4}  \frac{ \alpha^{3/2}} { \tilde K  \ K_{zz} ^{1/2}} \frac{1}{\alpha^3}
 \int   \frac{d^3 x}{(2 \pi)^3}
\frac{ \left[ (  \delta^3 r^6 \cos (6 \phi) - 3 \delta^2 \kappa r^5 z \sin (3 \phi) - \kappa^3 r^3 z^3 \sin (3 \phi) \right] }
{\left[ (1 + R^2)^2 - \delta^2 r^4 - \kappa^2 r^2 z^2 - 2 \delta \kappa r^3 z \sin(3 \phi) \right]^3}
\end{displaymath}
\end{widetext}
where $\delta \equiv ( \frac{K_{23}}{2 \tilde K})$ and $\kappa =  K'/ (K_{zz} \tilde K)^{1/2}. $,
and we have defined $r$ and $\phi$ by $(x,y) = r(\cos(\phi)), \sin(\phi))$.  We have also dropped terms
such as $\sin \phi$ and $\cos (2 \phi)$ in the numerator which vanish after integration. 
We see that $b$
is finite only when $K_{23}$ and $K'$ are both finite, and for small $K_{23}$ and $K'$, proportional to $ K_{23}^3 K'{}^2$.
Thus $b$ vanishes if the system has D$_{6h}$ symmetry. 
  In the same spirit as the approximations taken above,
the parametric dependences  of $b$ can be estimated as
\begin{equation}\label{b-est}
b \sim  - T  \frac{ \alpha^{3/2}} { \tilde K  \ K_{zz} ^{1/2}} \frac{1}{\alpha^3} \left[ \left(\frac{K_{23}}{\tilde K} \right)^3  \frac{( K')^2}{K_{zz} \tilde K} \right] 
\end{equation}
For simplicity of the presentation, we shall not display the numerical coefficient, which is found to be $33/ (4 \times 4096 \pi)$.  This small coefficient is again due to
the angular averages and high powers of $k$'s in the numerator of eq (\ref{b}), similar to what we have encountered in the estimation eq (\ref{a2}) for the second contribution to $a$. 
Correspondingly, 
\begin{equation}
\frac{ g_2 b^2}{c'} \sim  \frac{g_2} { 2 g_1 + g_2 } \left[ \left(\frac{K_{23}}{\tilde K} \right)^3  \frac{( K')^2}{K_{zz} \tilde K} \right]^2
\end{equation}
Thus $g_2 b^2$ is expected to give only a small contribution to $\tilde c$, especially if $g_2$ is small compared with $g_1$ or
when $K'$ is small etc.

We now estimate $\vert \vec h \vert$ at the first order transition at $T_1^*$ and compare it with $\alpha (T_1^*)$. 
We first note that $\alpha(T_1^*) > \alpha(T_1)$ since $T_1^* > T_1$. 
  Using eq (\ref{b-est}) and (\ref{c'1}) and (\ref{T1}), we get
\begin{equation} \label{hvalpha}
\frac{ \vert b \vert }{ c' \alpha(T_1)} \sim \frac{ \vert g_2 \vert} { (2 g_1 + g_2 )} \left[ \left(\frac{K_{23}}{\tilde K} \right)^3  \frac{( K')^2}{K_{zz} \tilde K} \right]
\end{equation}
with again  an expected small numerical factors implicit.  If we consider 
$ \vert b \vert /  [ \tilde c \alpha(T_1)] $, instead then a generous estimate would be to replace the 
term $( 2 g_1 + g_2)$ by $ ( 2 g_1 - \frac{3}{2} \vert g_2 \vert)$ as explained below eq (\ref{T10}).
Hence unless $\vert g_2 \vert \approx g_1$ and with very special circumstances for the gradient coefficients,
 we have $\alpha > \vert \vec h \vert$ at $T_1^*$, and the
superconducting order parameter nucleates only at a lower temperature, as sketched in Fig \ref{fig1}. 

Lastly we estimate $T_1^* - T_1$.    This is
\begin{equation}
T_1^* - T_1 \approx \frac{b^2} {\tilde c \left( \frac{ \partial \tilde a}{\partial T} \right) } 
\end{equation}
Note that $\frac{ \partial \tilde a}{\partial T}  = \frac{ \partial a}{\partial T}$.  If we again replace $\tilde c$ by $c'$, 
we get the estimate 
\begin{equation}
T_1^* - T_1 \sim  \frac{ \vert g_2 \vert} { (2 g_1 + g_2 ) } \left[   \frac{ ( \vert g_2 \vert T_0 )^2 } { \alpha' \tilde K^2 K_{zz} } \right]
\end{equation}
with a small coefficient due to $b$ implicit. Note again the appearance of a Ginzburg-like parameter on the right and compare this with eq (\ref{T10}).


\begin{thebibliography}{50}
\bibitem{footnote}  This condition is necessary but not sufficient.  For additional discussions, see, e.g., \cite{Barash}
\bibitem{g1h} It can be shown that all coefficients of $\vec h^{2N}$ with  $N  \ge 2$ obtained this way from eq (\ref{expand}) are positive.
\bibitem{reflect}  The fact that ${\rm sgn} h_x$ is independent of ${\rm sgn} K'$ can be understood as follows.
  A crystal with D$_{3d}$ has no $z \to -z$ reflection symmetry.  Under this reflection, the original crystal
would turn into a new one with $K_{23}$ remaining the same whereas $K'$ changes sign.  Under this reflection however,
the direction of the vestigial order (which is related to an in-plane distortion) hence $h_x$ does not change.
\bibitem{noteVR} \cite{VanderbosR} wrote the quartic terms of the free energy in a form similar to our
eq (\ref{Hintb}), thus their $B_1$ and $B_2$ are proportional to our $\beta_{12}$ and $-\beta_2$ respectively. 
The sign of our $g_2$ is therefore the same as that of $B_1 - B_2$ in the Suppl. Mat. of \cite{VanderbosR}.
Many different models are discussed there, but it seems to us that this quantity  in \cite{VanderbosR} is always non-negative.
\end{thebibliography}
\end{document}